\documentclass{sig-alternate-05-2015}
\newfont{\mycrnotice}{ptmr8t at 7pt}
\newfont{\myconfname}{ptmri8t at 7pt}
\let\crnotice\mycrnotice%
\let\confname\myconfname%

\CopyrightYear{2016}
\setcopyright{acmcopyright}
\conferenceinfo{WebSci '16,}{May 22-25, 2016, Hannover, Germany}
\isbn{978-1-4503-4208-7/16/05}\acmPrice{\$15.00}
\doi{http://dx.doi.org/10.1145/2908131.2908165}

\usepackage{amssymb}
\usepackage[font=small,format=plain,labelfont=bf,up,textfont=bf,up,justification=justified,singlelinecheck=false]{caption}
\usepackage{graphicx}
\usepackage{epstopdf}
\usepackage{epsfig,graphicx}
\usepackage{color}
\usepackage{url}
\usepackage{tabularx}
\usepackage{xfrac}
\usepackage{array}
\usepackage{subcaption}
\usepackage{multirow}
\usepackage{enumitem}
\usepackage{amsmath}
\usepackage{algorithm,algorithmic}
\usepackage{hhline}
\usepackage{booktabs}
\usepackage{booktabs}

\let\crnotice\mycrnotice%
\let\confname\myconfname%

\begin{document}
\clubpenalty=10000 
\widowpenalty = 10000
\newcommand{\tuan}[1]{{\textcolor{blue}{[\emph{\small Tuan: #1}]}}}
\newcommand{\khoi}[1]{{\textcolor{blue}{[\emph{\small Khoi: #1}]}}}

\newcommand\vpar{{\vspace*{1em}}}
\newcommand{\para}[1]{\noindent{\textbf{#1.}}}
\newcommand{\parai}[1]{\noindent{\textit{#1.}}}
\newcommand{\term}[1]{{\it {\small #1}}}

\newtheorem{mydef}{Definition}

\newcommand{\superscript}[1]{\ensuremath{^{\textrm{#1}}}}
\def\sharedaffiliation{\end{tabular}\newline\begin{tabular}{c}}
\def\ls{\superscript{1}}
\def\xls{\superscript{1,}}
\def\dfki{\superscript{2}}
\def\dk{\superscript{3}}



\addtolength{\itemsep}{-0.5in}
\setlength{\belowcaptionskip}{-10pt}

\title{Can We Find Documents in Web Archives without Knowing their Contents?}


\numberofauthors{1}
\author{
\alignauthor
Khoi Duy Vo, Tuan Tran, Tu Ngoc Nguyen, Xiaofei Zhu, Wolfgang Nejdl
\sharedaffiliation
       \affaddr{L3S Research Center / Leibniz Universit\"{a}t Hannover, Germany}\\
\vspace{5pt}
       \email{\{khoi,ttran,tunguyen,zhu,nejdl\}@L3S.de}
}

\maketitle
\begin{abstract}

  Recent advances of preservation technologies have led to an
  increasing number of Web archive systems and collections. These
  collections are valuable to explore the past of the Web, but their
  value can only be uncovered with effective access and exploration
  mechanisms. Ideal search and ranking methods must be robust to the
  high redundancy and the temporal noise of contents, as well as
  scalable to the huge amount of data archived. Despite several
  attempts in Web archive search, facilitating access to Web archive
  still remains a challenging problem.

  In this work, we conduct a first analysis on different ranking
  strategies that exploit
  evidences from metadata instead of the full content of
  documents. We perform a first study to compare the usefulness of
  non-content evidences to Web archive search, where the evidences are
  mined from the metadata of file headers, links and URL strings
  only. 
  Based on these findings, we propose a simple yet surprisingly
  effective learning model that combines multiple evidences to
  distinguish ``good'' from ``bad'' search results.  We conduct
  empirical experiments quantitatively as well as qualitatively to
  confirm the validity of our proposed method, as a first step towards
  better ranking in Web archives taking metadata into account.

\end{abstract}


 

\section{Introduction}
\label{sec:intro}

The Web exhibits rapid growth and dynamic changes of its structure and
contents, where ephemeral web pages are dominant, and content
disappearance is commonplace~\cite{ntoulas2004s}. A study by Toyoda et
al.~\cite{toyoda2003extracting} suggests that more than half of the
URLs disappeared or changed their location within one year. In order
to preserve parts of the Web for future generations, many Web
archiving initiatives are active, among which the Internet Archive is
the most prominent one. Web archive data, which can be up to petabytes
or more in size, consist of several snapshots of contents crawled from
the Web at different points in time. Such collections are valuable for
journalists, economists, historians, social scientists and others to
explore the past, but their values can only be uncovered with an
effective access and exploration mechanism. Such a mechanism should
rank documents not only by their content relevance to an information
need, as in the case of the actual Web, but also by their long-term
preservation values reflected in the archive. Furthermore, Web
archives are characterized by the high redundancy of contents, the
imbalance of revisions due to selective crawling
strategies~\cite{huurdeman2014finding}, and thus call for different
ranking approaches.

Searching in Web archives has drawn increasing attention in recent
years.
Early attempts focus on efficiently indexing the full contents of
documents, with optimization tailored to special types of
queries~\cite{berberich2007time}. Other recent work suggests that
non-content evidences such as timestamp~\cite{nguyen2015time},
hyperlinks~\cite{dai2010freshness}, or anchor
texts~\cite{dai2010mining,huurdeman2014finding} can be used to improve
ranking performance. For example, the anchor texts of hyperlinks, when
aggregated by the target page,
can well represent the collective perception of the document
essence~\cite{craswell2001effective}, in contrast to the actual
content of the document which reflects the author's intent. Similarly,
authority evidence of a web page such as its PageRank score can
complement the relevance of the page with its importance or
freshness~\cite{nguyen2015time}. Despite several approaches having
been proposed and developed, no optimal search strategy for Web
archives has been developed yet. Furthermore, while the existing work
suggests the benefit of different individual evidences, no study has
compared the impacts of these evidences in a unified ranking
framework, their contribution as well as disadvantages when put
altogether.

In this paper, we follow the above school of thought, and argue that
using other information than the document contents can help finding
relevant and important documents with adequate performance, while
avoiding the high cost of fully indexing the Web archive. Furthermore,
we provide the first study to compare the usefulness of non-content
evidences in web archive search, where the evidences are mined from
the document metadata such as URL strings, file headers, and from the
links such as anchor texts and the linking statistics. We conduct
experiments with an entity search scenario in mind, i.e. finding
documents related to an entity from the Web archive, and focus on
entities with low ambiguity to reduce the effect of spurious
results. In summary, we investigate the following research questions:

\begin{itemize}
\item \textbf{RQ1}: How useful are non-content features of documents
  for finding relevant documents in a Web archive?
\item \textbf{RQ2}: Does the combination of multiple features improve
  the performance over individual ones?
\end{itemize}

We explore the above questions on subsets of the Internet Archive
dataset, with our experiments focusing on the \textsf{.de} domain. We
investigate the problem of exploration in the Web archive, where the
queries are unambiguous entities identifiable from Wikipedia, and the
documents are web pages with several revisions captured in the
archive\footnote{In this paper, we will use ``document'' and ``web
  page'' interchangeably.}.

We examine several features derived from different sources of the
documents, without processing the full text of documents. While the
results confirm our assumption that non-content evidences are valuable
resources and deserve more attention, they also give interesting
insights into the influences of features in different settings. Based
on the findings of this analysis, we propose a simple yet effective
ranking model to combine multiple evidences for distinguishing
``good'' from ``bad'' search results in a Web archive, without knowing
their full contents, and a scalable approach to obtain the training
data for our learning models without much human effort.

The remainder of the paper is organised as follows. After the
discussion of related work in Section \ref{sec:rw}, in Section
\ref{sec:prep} we describe the dataset used in our study, the
characteristics of the dataset as well as some first exploratory
analyses. Section \ref{sec:analysis} reports the analysis in Web
archive search using non-content evidences. In Section
\ref{sec:ranking}, we discuss our ranking model based on non-content
features, and show that it is indeed able to distinguish ``good'' from
``bad'' search results. Finally, in Section \ref{sec:conclusion}, we
close with a discussion of future work and conclusions.


\section{Related Work}
\label{sec:rw}

Related work can be categorised into three groups. The first group
consists of work on building infrastructure for indexing and accessing
Web archives. The second group consists of work on exploiting
non-content features in information retrieval. As there exists a vast
amount of work in this area, we focus on existing work on retrieving
documents using anchor texts, as this provides several interesting
questions for our analysis. The third group of work consists of
ranking approaches in Web archives.

\subsection{Indexing and Searching Web Archives}

Web Archives are used to preserve digital heritage and the web for
future use~\cite{masanes2006web}. One of the major issue in this
context is how to provide infrastructures for indexing, searching, and
access to these often huge archives \cite{stack2006full}. This lead to
a bulk of technical difficulties which have to be solved.
Stack~\cite{stack2006full} proposed a modification of Nutch to enable web
archive indexing and full-text search on larger-scale collections,
which increase the capacity of Nutch from under 100M to higher
amounts. Lin et al. \cite{lin2014infrastructure} provide a scalable
tool using Map-Reduce paradigm to parse data from and to the index.

\subsection{Exploiting Anchor Texts for Retrieval}
\label{subsec:anchor}

Anchor text of links in documents has been widely adopted as promising
information in the context of Web search.
A number of studies find that anchor text is useful for information
retrieval when queries are navigational~\cite{craswell2001effective,
  fujii2008}, as well as for ad hoc information search~\cite{koolen2010}.
In \cite{craswell2001effective}, Craswell et al. aggregated anchor
texts for a target page and used them as surrogate documents for
finding sites. In~\cite{fujii2008}, Fujii et al. studied the
effectiveness of content-based or anchor-based retrieval methods for
different query types. Both~\cite{craswell2001effective,fujii2008}
concluded that incorporating anchor text can significantly improve the
retrieval effectiveness for navigational queries. Koolen et
al. \cite{koolen2010} further investigated the importance of anchor
text for ad hoc information search and showed that anchor text based
methods can significantly improve retrieval effectiveness.



In the context of Web archive search, anchor texts can help in
additional interesting ways. For example, they can be used to address
the incompleteness of the archive collections, and help retrieving
\emph{unarchived} Web pages, or Web pages that were once present in
the Web but not preserved in the archives.
Klein and Nelson~\cite{klein2014} leveraged the top-$n$ words of link
anchors to extract lexical signatures of such missing webpages, which
were then used to retrieve alternative URLs for these missing
webpages. Similarly, Huurdeman et al.~\cite{huurdeman2014finding}
conducted research to use links and anchors in the crawled webpages to
recover unarchived webpages.
Other research work combined anchor texts with additional
information. Some researchers aggregated anchor texts from all pages
that link to a page in the same domain~\cite{metzler2009building}, or
the same website~\cite{kleinberg1999,kraaij2002} as the target page.
Regarding time, historical trends of anchor texts have also been
investigated for estimating anchor text importance. For instance, Dai
et al.~\cite{dai2010mining} differentiated pages with different
in-link creation rates, and concluded that ranking performance can be
improved via taking into account in-link creation rates. Nguyen et
al.~\cite{nguyen2015mining} study the problem of mining temporal
subtopics in the Web Archive. They propose to mine the relevant time
of temporal subtopics by leveraging the trending behaviour of the
corresponding anchor texts.

In this paper, we focus on a different scenario, an exploratory search
scenario, and find that anchor texts alone do not provide an optimal
solution. We incorporate anchor text and other non-content features,
without involving content of webpages, and show that this combination
is better able to distinguish between good and bad search
results. 

\subsection{Ranking Models in Web Archives}

There is a number of approaches addressing the issue of ranking in the
Web archives, which can be classified into three main lines. The first
approach is based on full content indexing, suggested in the early
work by Berberich et al. \cite{berberich2010language}. This approach
takes into account temporal expressions and integrates them into a
language model retrieval framework. However, the method only works for
limited types of queries. The second approach consists of graph-based
methods, exploiting the hyperlinks in the documents of the archive. In
\cite{dai2010freshness}, Dai et al. leverage features from historical
author activities and propose to consider the authority over multiple
web snapshots at different time points. They modify the traditional
link-based web ranking algorithms by further incorporating web page
freshness over time from page and in-link activity. Nguyen et
al. \cite{nguyen2015time} attempt to discover content that can cover
most interesting time periods for a given topic. To this end, they
design a novel graph-based model by integrating relevance, temporal
authority, diversity and time in a unified framework. A third line of
research is represented by the learning approach in Costa et
al. \cite{costa2014learning}. They assume that closer time periods are
more likely to hold similar web characteristics. Based on this
assumption, they propose a novel temporal-dependent ranking framework
which exploits the variance of web characteristics over time. In
contrast to this work, our work targets ranking documents and not
revisions, and focuses on non-content features, to provide a more
light-weight ranking framework.

\section{Web Archive Data}
\label{sec:prep}

\subsection{Dataset} 

In this work, we use subsets of Web archive data provided by the
Internet Archive from September 1996 to the end of December 2014. We
focus our study on web pages with the \textsf{.de} top-level
domains. 
This collection consists of 762,008 archive files, including 124,743
\textsf{.arc} files and 637,265 \textsf{.warc} files, which comprise
55.6TB in volume\footnote{All collection files are compressed in gzip format}. It contains 1,434,118,956 URLs and 141,835,258,519
revisions, belonging to 9,880,121 of domains and 702,398,802 core URLs
(i.e., URL obtained after excluding query strings).
Revisions are not evenly distributed over URLs, but instead heavily
biased to a small portion of domains, due to a number of selective
crawling strategies \cite{huurdeman2014finding}. This is known as the
incompleteness problem, which is ubiquitous not only in the Internet
Archive dataset, but in all Web archives
\cite{huurdeman2014finding}. We take this characteristic into account
in our analysis in the next sections.

In this paper, we focus on investigating the usefulness of document
metadata in Web archives without analysing the full contents. The
metadata come from the following major sources: From the URL of the
document, from the header of the archive files (\textsf{.arc} and
\textsf{.warc} formats), and from the hyperlinks. As for the URL, we
tokenize the string into different words, use a fixed set of defined
delimiters (e.g. ``\texttt{/}'', ``\textsf{.}'', ``\textsf{-}'',
``\textsf{\_}'' ). The hyperlink information requires slightly more
work to be extracted, we describe this separately in the following.

\subsection{Preprocessing}

\vpar
\para{Web Archive Graphs}
Hyperlink information has been proven to be a powerful indicator for
searching in Web archives \cite{nguyen2015time,
  dai2010freshness}. Inspired by previous work, we extracted the link
structure of the Web archive documents in our set. There are
305,726,983,071 links between revisions, with 14 link types (shown in
Table ~\ref{table:linktype}). Most of link types are links to
background or images in documents. We select only links to content
pages, identified by the tag \textsf{<a>}. There are 161,728,553,318
links after removing noisy pages. The existing links contain anchor
texts to a document revision in the archive, with approximately 2
links per revision in average.

\begin{table}[h]
    \scriptsize
    \centering
    \begin{tabularx}{\columnwidth}{X|l}
        \hline
        HTML Tag pattern & Destination type\\
        \hline
        A\@/href & \textsf{Either a hyperlink or an anchor}\\
        IMG\@/src & \textsf{Image source}\\
        AREA\@/href & \textsf{Area inside an image-map}\\
        EMBED\@/src & \textsf{Embeded object}\\
        FRAME\@/src & \textsf{Frame source}\\
        INPUT\@/src & \textsf{Input source}\\
        IFRAME\@/src & \textsf{Iframe source}\\
        FORM\@/action & \textsf{Form processing URL}\\
        TD\@/background & \textsf{Table cell background image}\\
        TR\@/background & \textsf{Table row background image}\\
        BODY\@/background & \textsf{Page background image}\\
        OBJECT\@/codebase & \textsf{Code of embeded object}\\
        TABLE\@/background & \textsf{Table background image}\\
        FB:LOGIN-BUTTON\@/background & \textsf{(Facebook
                  login) button background image}\\ 
        \hline
    \end{tabularx}%
    \captionsetup{justification=centering}
    \caption{List of link types and their patterns, as used to
          extract from the Internet Archive data} 
    \label{table:linktype}%
\end{table}%
%


\vpar
\para{Web Archive Anchor Texts} One special aspect of a hyperlink is
the anchor text, which is a visible text snippet used to label the
links. To some extent, anchor texts encode the testimony of the users
to the target, and when put together collectively, can represent a
surrogate of the document. We extracted 161,728,553,318 distinct
anchors text links from the hyperlinks, with 118 links per document on
average.

\begin{figure*}[ht]
    \centering
    \includegraphics[width=2\columnwidth]{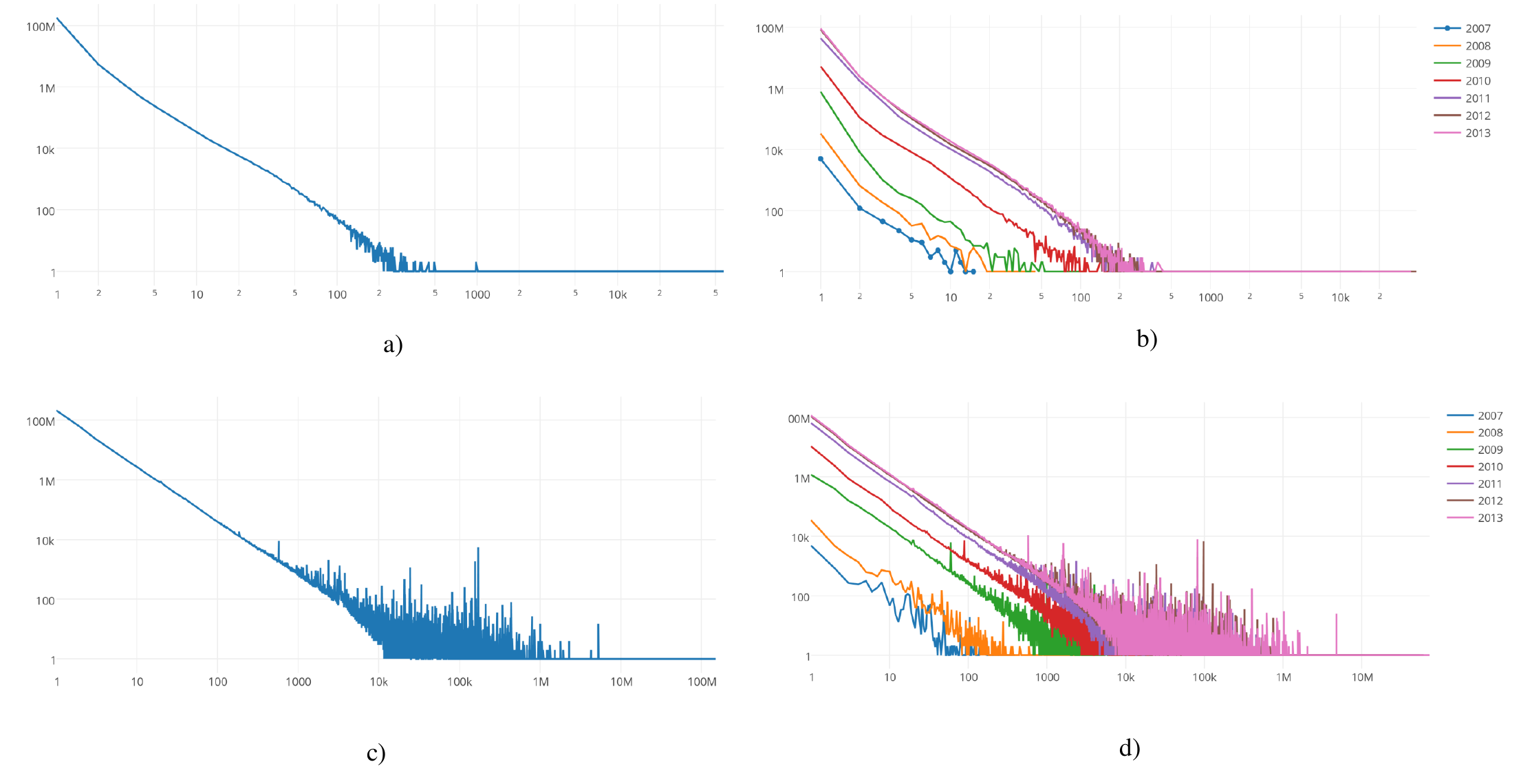}
    \caption{Distribution of Anchor Texts in \textsf{.de} domains
          of Internet Archive. Y-axes correspond to the number of
          anchor texts, X-axes are the count of target urls per each
          anchor text. All axes are in log scales. a) and b) are
          distributions for the 350 biggest domains, judged by the number of
          Web pages, c) and d) are distribution over all domains. b)
          and d) are evolutions of anchor text distribution over years
          in the archive.}
    \label{fig:anchor3}
\end{figure*}

\vpar
\para{Indexing} Inspired by early work on anchor text-based search
\cite{craswell2001effective}, we represent a web document by
concatenating all of their anchor texts pointing to this document, and
utilize a full-text index to index this document ``anchor
representation''. We use
ElasticSearch~\footnote{\url{https://www.elastic.co/}} to achieve
indexing capacity at large scale.
One subtle issue is that in Web archives, a link can be repeated
multiple times due to crawling strategies. For such links, we decide
to perform two strategies: (1) keep one unique anchor text for a
source-destination pair for each revision, and (2) keep all anchor
texts across revisions, even it is possible (and likely) that such
revisions are identical. This allows us to exploit the accumulated
statistics of anchor texts over time, and to analyse the effect of
time on the anchoring behaviour\footnote{In strategy (2), we truncate
  the anchor texts of 26 URLs of which the length exceeds the storage
  limit of ElasticSearch.}.

There are 26,443,384,902 destination URLs for all links in web
archive. However, we only make use of the 
ones which are archived in our dataset, resulting in 990,031,302
corresponding documents. Among these, there are 100,047,693 documents
that have no anchor text at all, which we remove, reducing the number
of retrievable documents to 889,983,609. This data is indexed by
ElasticSearch as mentioned above. The resulting documents serve as
potential search results from the archive.
\subsection{Anchor Texts Distribution}

As mentioned above, anchor texts have been widely used as a useful
meta-data besides full contents to virtually represent the (target)
document, with a rich body of related work (Section
\ref{subsec:anchor})
In this paper, we investigate this further, by analysing the
distribution of anchor texts in the context of Web archives.

Figures \ref{fig:anchor3}a-d show the distribution of anchor texts
measured by the number of destination Web pages in log-log
scales. From these figures, a heavy-tail distribution of anchor text
usage can be observed. Another observation is that anchor texts follow
power law distribution quite strictly in the head part, while the tail
part exhibits high noise for popular anchor texts. The degree of noise
increases over the years, as shown in the Figure
\ref{fig:anchor3}d. In this figure, we group the distribution by
projecting timestamps of source Web pages for each hyperlinks,
grouping them on a yearly scale, drawing each distributions for each
year, and putting them in one timeline from 2007 to 2013. From this
timeline, we can see the continuous increase of noise in the tails
over the years. We believe that this is mainly due to the imbalanced
number of snapshots of websites according to the crawling strategies,
which results in a bias in the sampling of domains in the archive
(i.e., not all Web pages in the domain are crawled, or are crawled at
different frequency).

To verify this assumption, in Figures \ref{fig:anchor3}a and
\ref{fig:anchor3}b, we analyse and show the distributions of the top
$350$ biggest domains, judged by the number of member Web pages
preserved under the domain. For these domains, we observe that the
crawling policy is more consistent over time, i.e.  Web pages are all
aggressively and comprehensively crawled regardless of their active
time periods. This results in the balanced number of revisions crawled
for both rare and popular anchor texts drawn from these domains. In
the figures, the tail becomes much less
noisy. 
This finding is interesting, as it reveals that the revision counts
(driven by the crawling policies) have an implication on distribution
of anchor texts, making anchor texts different from those normally
drawn from the contents of the Web pages, suggesting that they should
not be exploited in isolation, but rather in combination with other
evidences such as the size and crawling frequency of domains. In the
next sections, we look further into the correlation of features and
the quality of retrieved documents.

\section{Analysis on Web Archive Search}
\label{sec:analysis}

As we are interested in exploiting non-content information for search,
we investigate the two research questions described in
Section~\ref{sec:intro}. For question 1, we focus on \emph{entity
  search} scenarios, described below, as the anchor texts for entities
are more intuitive and interpretable, and also less noisy. For
question 2, we analyse the correlation between relevance of documents
and different non-content features.

\subsection{Entity Queries}

\para{Analysis Setup} In order to avoid spurious effects of ambiguous
queries in Web archive search, we limit our queries to entities
identifiable by a Wikipedia page. We choose Wikipedia pages that are
not list pages, disambiguation pages, and contain no commas and round
brackets which indicate potential ambiguity of entity names
\footnote{\url{https://en.wikipedia.org/wiki/Wikipedia:Article_titles}}. Our
query set consists of $216$ queries, chosen from entities with both
high and low popularity\footnote{The full list of queries are omitted
 due to space limit, and available upon request from
 khoi@L3S.de}. Specifically, we use the page view counts (extracted
using Hedera~\cite{tran2014hedera}) of a Wikipedia page as proxy for
the corresponding entity popularities, partition the view counts into
three buckets (Top, Middle, Bottom), and randomly choose entities in
each partition to build the sample. We also group entities into
different types to facilitate high-level analysis. We first used
DBpedia knowledge base \cite{auer2007dbpedia} to resolve the types of
the sampled entities, and manually mapped each entity into one most
plausible, coarse-grained type. Table \ref{table:queries} lists the
types studied in our work.
We acknowledge that the list is still primitive, and leave a deeper
analysis with a more advanced taxonomy for the future (for example,
multiple classes per entity).

\begin{table}[h]
\small
  \centering
    \begin{tabular}{rl}
    \toprule
    \textbf{Type} & \textbf{Examples} \\
    \midrule
    Politician & \textit{Barack Obama}, \textit{Angela Merkel} \\
    Scientist & \textit{Max Planck}, \textit{Josef Meixner} \\
    Artist & \textit{Michael Jackson}, \textit{Scorpions} \\
    Sport player & \textit{Andreas Nilsson}, \textit{Franz Beckenbauer} \\
    Author & \textit{Franz Hoellering}, \textit{Artur Kaps} \\
    Entrepreneur & \textit{Enzo Ferrari}, \textit{Artur Mest} \\
    Organisation & \textit{Maharashtra Film Company} \\
    Product & \textit{Chevrolet Vega}, \textit{Apple Macintosh} \\
    Location & \textit{Volken}, \textit{Larrabee State Park} \\
    Works & \textit{Volken}, \textit{Alien Trespass} \\
    Event & \textit{FIS-Ladies-Winter-Tournee 2011} \\
    Biology & \textit{Rote Mangrove}, \textit{Ortolan} \\
    Music & \textit{Appenzeller Streichmusik} \\
    Astrology & \textit{CloudSat}, \textit{Abell 262} \\
    Abstract Concept & \textit{Aequat Causa Effectum}\\
    \bottomrule
    \end{tabular}%
  \captionsetup{justification=centering}
  \caption{List of Entity types}
  \label{table:queries}%
\end{table}%



\vpar
\para{Example Analysis} Figure \ref{fig:merkel} shows the query
``\textsf{Angela Merkel}'' as covered in the Web (worldwide and in
Germany), and as covered in our \textsf{.de} web archive. The Web
coverage timelines are derived from Google Trend, while for the
\textsf{.de} web archive, we derive them from our anchor text index
described in Section \ref{sec:prep}. All values in the timelines are
normalized to $[0,1]$ by dividing to the max value.

The first observation from these timelines is that query coverage on
the Web follows bursty patterns, with big spikes reflecting important
events, quite synchronized between German and international
websites. For instance, the first spikes during the end of 2005
correspond to the event of Angela Merkel starting her duty as the
chancellor of Germany. As for the archived anchor dataset, the
trendings are only visible during the period of big events observed
from the actual Web, i.e. the end of 2005 (her first term as the
chancellor) and the period of 2011-2012 (active period of the Eurozone
crisis). For many other medium size events, we do not observe peaks,
and the ratio of coverage stays relatively stable. We believe that the
main cause for this asynchrony stems from the fact that the purpose of
putting an anchor text in a hyperlink from one content to another is
to endorse the destination content, or to establish the contextual
relevance between the two contents. This is different from queries on
the actual Web, which directly encodes the user information need. The
implication is that any existing work that relies purely on anchor
text as the source of retrieval can only gain adequate performance on
certain sets of queries such as finding specific URL
\cite{craswell2001effective}, or searching for highly debated topics
\cite{nguyen2015time}. For other types of information need, combination
with other evidences is necessary.


\subsection{Document Metadata and Search}


\para{Analysis Setup}
In this section, we investigate the influence of different non-content
evidences on Web archive search. To start with, we choose the three
main sources of evidences for the preliminary analysis:

1. \noindent \textit{URL}: We choose to investigate the depth of the
document URL, i.e., the level of sub-domains or \emph{directory} in
which the corresponding web page resides in the Web server. The lower
URL depth corresponds to more general web pages such as home pages,
while higher depth indicates that the document represented by the web
page is likely to mention more specific information (such as product
detail in an e-commerce site).

2. \noindent \textit{Anchor Text}: We count the frequency of the query
as appeared in the anchor text representation of the document (see
Section \ref{sec:prep}, Indexing) as another evidence.

3. \noindent \textit{Revision Number}: We count how many times a web
page has been crawled. To some extent, this reflects the importance of
the web page, as we observe popular and highly dynamic web pages to be
captured more often than others.

\begin{figure}[ht]
\centering
\includegraphics[width=\columnwidth]{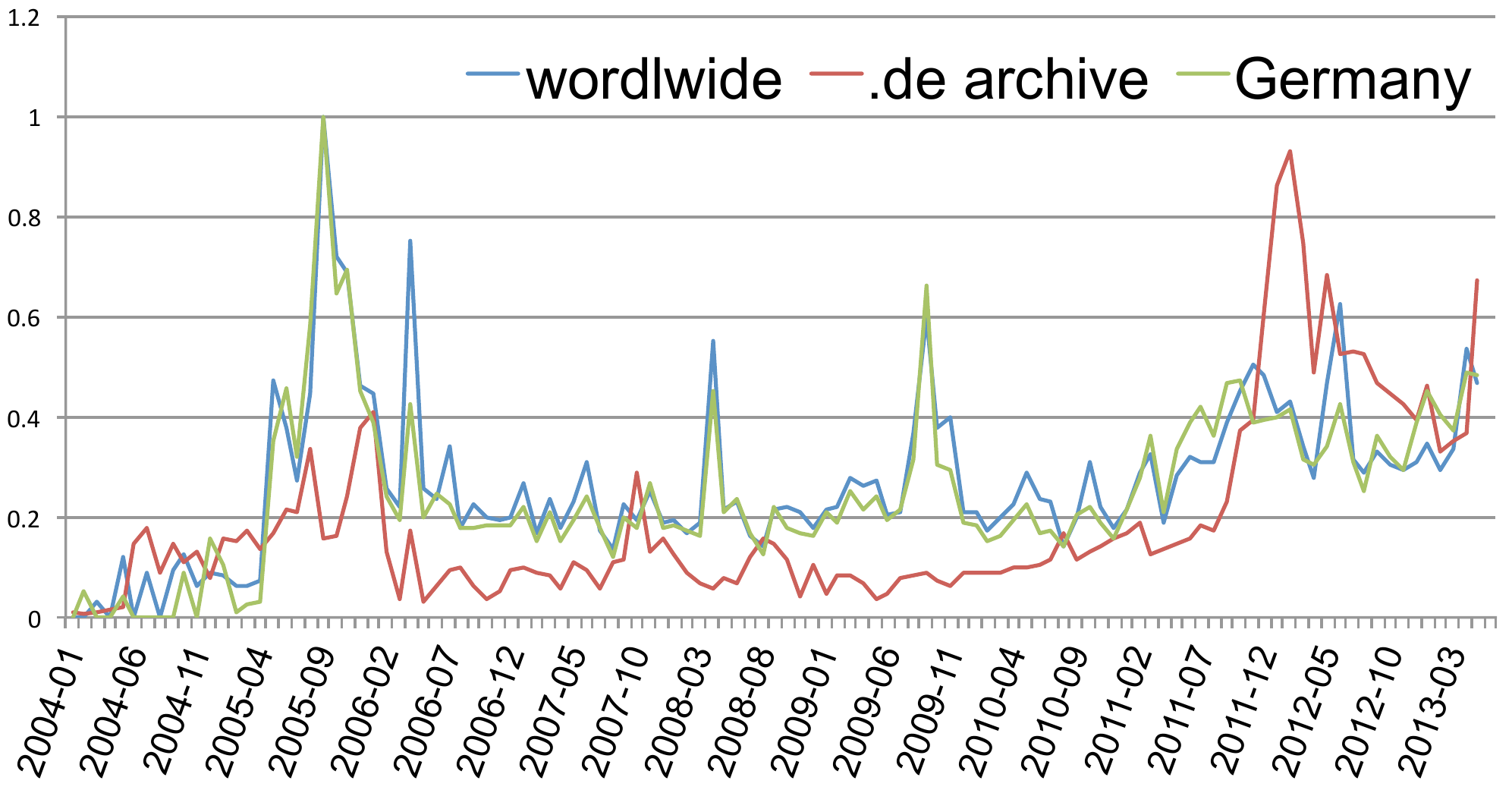}
\caption{The timeline of coverage of the query ``Angela Merkel'' from
  the Web (worldwide and in Germany), as compared to the timeline of
  coverage in the \textsf{.de} domain in our Internet Archive
  dataset.}
\label{fig:merkel}
\end{figure}

\begin{figure*}[htb]
\centering
\includegraphics[width=2\columnwidth]{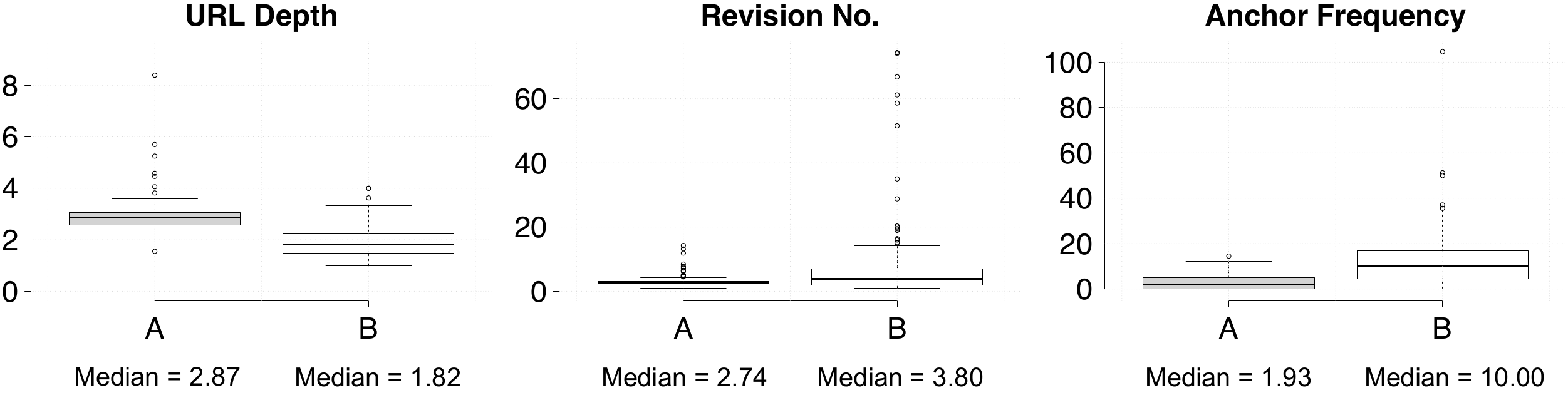}
\caption{Distributions of the three evidences in all results returned
  by the archive index (A), and in top results returned by Bing
  (B). Each point corresponds to an average value of the evidence
  derived from all results of a single query} 
\label{fig:medians}
\end{figure*}

\begin{figure*}[htb]
\centering
\includegraphics[width=2\columnwidth]{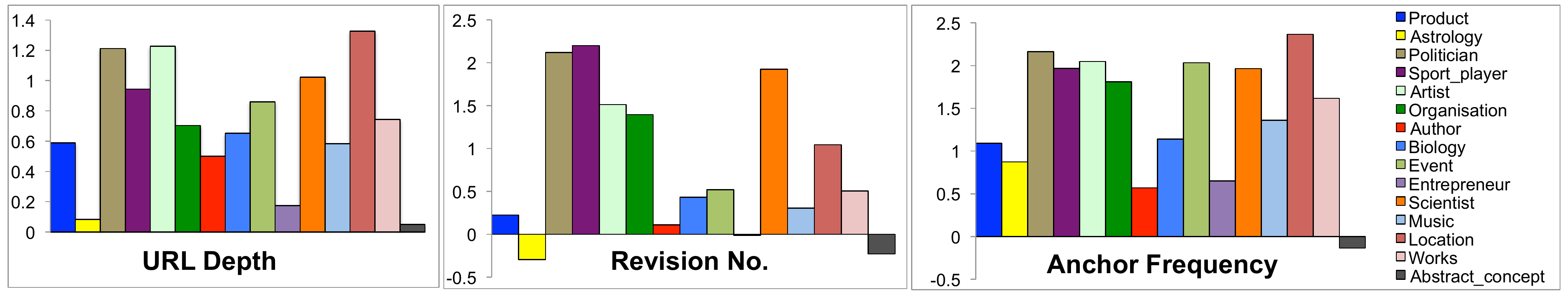}
\caption{Distributions of the three evidences from URL, revision count
  and anchor. Each column corresponds to an average value of the
  evidence for one query type (derived from top results by Bing). The
  y-axes are evidence scores in log scale} 
\label{fig:3dims}
\end{figure*}

\vpar
\para{Bing Search Engine as relevance proxy} For each entity query, we
query our index to get all documents that contain the query either as
part of anchor text, or in the URL string of the corresponding web
page. We call this dataset $A$. This dataset contains $1,276,900$ URLs
per query in average (median value is $37199$ URLs per query). We also
issue the query to the Bing search engine and fetch the URLs from the
top-$100$ results that overlap with our results from the index. We
call this dataset $B$. The intuition for this setting is that we
assume top results returned by Bing are more likely to be of higher
quality and relevance to the query of interest. To somewhat reduce the
bias introduced by time, we issued the query to Bing at different
points in time over the last half year, 
and merge the top-100 URLs of all results.

\vpar
\para{Analysis}
Figure \ref{fig:medians} shows the distributions of the averaged
evidence scores per query, as derived from result set $A$ and $B$. It
is calculated by obtaining retrieved documents for each entity query
(either returned by Bing or by the local ElasticSearch index),
calculating the corresponding feature values for each document and
then getting the average. We observe that documents returned on the
Bing first page tend to have lower depth of URL strings as compared to
documents in general (median $1.82$ compared to $2.87$), as well as
higher number of revisions ($3.80$ compared to $2.74$), and appear in
more links where the queries are used as the anchor texts ($10.00$
compared to $1.93$). On the other hand, these evidence scores of top
results exhibit higher degree of diversity, reflected by the wider
quartile windows in all three (for the Anchor Text evidence, the first
and third quartiles of top result are $[4.56, 16.81]$ compared to
$[0.00, 5.15]$ for all results).

In Figure \ref{fig:3dims}, we further investigate the correlation of
this evidence diversity and the entity types. The scores, which are
obtained in the same way as for Figure \ref{fig:medians}, are presented
in log scales, thus include some minus values (for instance, the query of
``Abstract Concept'' entity type appear only $0.59$ times on average
in the top URLs returned by Bing). For different entity types, we
observe a big difference in the scores. For example,
queries about politicians or sport players tend to have results with
deep URL strings, as they often appear in news or other well-organised
websites. They are also captured more often than others, reflected in
higher average revision counts. On the other hand, queries about
authors (play writers, novelists) have the least crawled results,
mainly because many of them have static contents. As for the URL depth
and anchor frequency, the highest average scores are for queries about
location. Analysing deeper, we observe that the location terms appear
much more frequently in the result web pages, including boilerplated
parts of the pages such as disclaimers, ads, headers, etc. This
preliminary analysis suggests that in general, any search system that
deals with entity queries must take into account different dimensions,
for example the inherent semantics of the entity
types.

\section{Ranking by Multiple Evidences}
\label{sec:ranking}

Based on our previous analysis, in this section, we investigate the
effectiveness of combining multiple evidences from metadata to
distinguish good from bad search results from the Web archive. We
extract different features and rely on learning-to-rank models to
provide a unified ranking
score. 

\subsection{Features}

Table \ref{tb:features} lists the set of features investigated in our
ranking models. In the following we explain some selected features.

\vpar
\noindent \textbf{NewsURL}: This feature estimates the newsworthy of a
content via its domain. We curate from the Wikipedia page
``Liste\_deutscher\_Zeitungen'' and from the web page ``Paper
Boy''\footnote{\url{http://www.thepaperboy.com/germany/newspapers/country.cfm}},
among other sources, to obtain the list of 507 German news sites.

\begin{table*}[ht]
{\scriptsize
\centering
\renewcommand{\arraystretch}{1.3}
\begin{tabular}{|c|c|l|}
 \hline
 \textbf{Source} & \textbf{Name} & \textbf{Description} \\
 \hline
 \multirow{6}{*}{URL} & URLDepth & The depth of the document URL \\
 \cline{2-3}
 & QueryString & Whether the document URL contains query strings \\
 \cline{2-3}
 & SearchWord & \parbox{12cm}{Appearance of searching words (``such'',``suchergebnis'', ``query='', etc.) in the document URL} \\[5pt]
 \cline{2-3}
 & QueryInURL & Frequency of query terms in the document tokenized URL\\
 \cline{2-3}
 & NewsURL & Whether the document URL belongs to a news domain \\
 \cline{2-3}
 & WikipediaURL & Whether the URL has been cited in the Wikipedia page of the queried entity \\  
 \hline
  \multirow{3}{*}{Links} & InlinkCount & The number of incoming links pointing to the document in the Web archive \\
 \cline{2-3}
 & PageRankCore & PageRank score of the document in the hyperlink graph\\
 \cline{2-3}
 & PageRankDomain & \parbox{12cm}{PageRank score of the domain in the hyperlink graph, where nodes are projected to domains} \\[5pt]
 \hline
 \multirow{3}{*}{Anchor Texts} & AnchorFreq & \parbox{12cm}{The fraction of incoming links pointing to the document in the Web archive, and having queries in their anchor texts} \\[5pt]
 \cline{2-3}
 & AnchorTimeSpans & \parbox{12cm}{Time spans of two consecutive timestamps of the anchor texts that are longer than 1 week} \\[5pt]
 \cline{2-3}
 & LuceneTf & Max value of Lucene term frequencies \cite{eslucene}\\
 \hline
 \multirow{7}{*}{Metadata} & DocLen & Document length counted by anchor texts \\
 \cline{2-3}
 & FieldNorm & Lucene field-length normalization score of anchor texts \cite{eslucene}\\
 \cline{2-3}
 & Idf & Inverse frequency of documents contain the query terms \cite{eslucene}\\
 \cline{2-3}
 & Revision & Number of revisions of the document \\
 \cline{2-3}
 & RevDuration & Count of time spans of two consecutive revisions that are longer than 1 week \\
 \cline{2-3}
 & DomainSize & Count of web pages having the same domain as the document \\
 \cline{2-3}
 & EntType & Type of entity represented by the query \\
\hline

\end{tabular}}
\caption{Selected features for analysing the ranking models, as
  grouped into different categories of evidence sources} 
\label{tb:features}
\end{table*}

\noindent \textbf{WikipediaURL}: This feature indicates the
credibility of the content via its domain. We assume that domains that
have more citations from the Wikipedia page of the entity query are
more topically relevant. For instance, many citations from the
Wikipedia page of \textsf{Michael Jackson} are from entertainment
websites, while in \textsf{Max Planck}'s Wikipedia page they are
mainly from scientific websites. We measure this feature by the number
of times a domain is cited in the Wikipedia page of the entity.

\noindent \textbf{AnchorTimeSpans}: Besides the anchor text frequency
feature (mentioned in Section \ref{sec:analysis}), we also consider
the temporal dimension of the links. The intuition is that in the Web
archive, the time when the anchor texts appear also reveals some
information about how important the target is. If two anchor texts
come from two links of revisions that have close timestamps, they
might not count as completely separate endorsement. For example,
if a Web page is crawled two times within a few hours, the contents
might be just identical, and therefore counting the anchor texts two
times will give false high endorsements to the target
documents. Hence, for this feature, we measure the distance between
any two consecutive timestamps of the anchor texts, and count the
number of time spans of which the distance is longer than $1$ week.

\noindent \textbf{RevDuration}: This feature relies on the frequency
of the crawlers to estimate the quality of the documents. If the
documents have many revisions that are crawled within a short time
period (e.g., a few days), it might have less authority than other
documents with revisions crawled in different years (the Web page are
still relevant to users, or are still referred to from other Web pages
posted later). For this feature, we measure the number of durations
between two consecutive revisions that are at least $1$ week long.

\noindent \textbf{PageRank}: We also calculate the authority scores of
the document by constructing the graphs of different links. The
PageRankCore scores are calculated on the graph of direct hyperlinks
between two web pages, and the PageRankDomain scores are calculated
based on links between domains.

\subsection{Experiment Setup}

\subsubsection{Training Data and Sampling} To study how the features
can be combined in a unified learning-to-rank model, we need the
relevance feedback for documents in the Web archive, or the training
data labels. Since there is no standard benchmark for this problem,
and since the focus of our work is to study the influence of features
coming from different evidences, in this first study, we choose two
pragmatic approaches for building the training data labels, described
subsequently. First, we choose a subset of 15 entity queries from the
testbed for our further experiments and case study analysis. As
judging all URLs returned in the dataset $A$ is infeasible, we decide
to evaluate on a small random sample of the results, constructed as
follows. For each feature dimension, we divide the results into three
partitions according to the order of their normalized scores. Then we
randomly pick $20$ to $50$ Web pages from each
partition. 
The sample is then obtained by pooling all Web pages from all
features. 
Our sampling algorithm makes sure there are overlaps between the
samples of each feature pair, so as to reduce the total size of the
pool. In practice, this results in $400$ sampled documents per query
in average. Then, two labeling strategies are applied:

\vpar
\noindent \textbf{Soft Labeling}: Similarly as the approach discussed
in Section \ref{sec:analysis}, we compare the results retrieved by our
index with the top-$100$ results returned by Bing. If found, the label
of the document is measured by the inverse of Bing rank, otherwise the
document is labeled zero. This enables us to exploit the order in Bing
search results for pairwise comparison in a learning-to-rank
model. The advantage of this approach is that we can easily scale up
training data construction in future work.

\vpar
\noindent \textbf{Manual Labeling}: Besides the automated approach
using Bing search results, we also experimented with manual labeling.
Four evaluators annotated the documents by checking their content in
the browser and in the Internet Archive \url{archive.org} site. Each
evaluator annotated the documents from the sample on a three value
scale: 0 means irrelevant, 1 means relevant but not important, and 2
means relevant and important documents. Here the notion of
``importance'' is defined by guiding the evaluator to think of the
document in a long term exploration scenario, and estimate, whether
the document contains lasting information about the queried entity or
not. For example, a biography page about Albert Einstein should be
labeled as 2, while an ad page about a product with Albert Einstein
portrait should be labeled as 1. The average Cohen's Kappa for the
evaluators' pairwise inter-agreement is $\kappa=0.75$, suggesting that
the assessment task has fair cognitive complexity.

\vpar \noindent \textbf{Adding Positive Examples}: To make sure the
training data have enough positive and negative labeled URLs of both
strategies, after pooling Web pages by features, we include all
results of our dataset $B$ into the sample. We acknowledge that the
sample has a strong effect on the results and models learnt, and plan
future experiments with more advanced sampling strategies and larger
samples. However, our current samples at least allows us to answer the
question how well we can distinguish between good and bad search
results, even if we cannot extrapolate these results to our full
dataset A, which contains many more potential search results.

\subsubsection{Ranking Methods}
\para{Baselines} We consider the following baselines: 1) The
BM25-based ranking score as returned from our anchor text-based index;
2) PageRank score; 3) Scores based on frequency of queries in the
document URL string (QueryInURL). Each of these baselines corresponds
to one source of evidence, discussed in the analyses in Section
\ref{sec:analysis}.

\vpar
\para{Learning Models} For the learning-to-rank method, we employ
Random Forests (RF)\cite{breiman2001random} as implemented in the
RankLib toolkit\cite{ranklib}. We use $5$-fold cross validation and
Normalized Discounted Cumulative Gain (NDCG) scores at top-$10$
results to optimize the parameters. RF models are trained on both
training data with soft labels derived from Bing (denoted RF-B) and
with manual labels (denoted RF-M). In each case, we use the labels of
the same strategy as ground truth for the evaluation (so RF-B is
evaluated against Bing results and RF-M against human feedback). As we
investigate in this first study the potential of using non-content
features to retrieve documents, we leave sophisticated settings of
cross-model and cross-label evaluations for the future.

\subsection{Empirical Results}
\label{sec:res1}
We use standard information retrieval metrics Precision (at different
cut-off threshold of 1 and 10), NDCG as well as MAP for the evaluation.

\begin{table}[htb]
\centering
\begin{tabular}{@{}ccccc@{}}
\toprule
      & P@1     &  P@10    & NDCG@10 & MAP\\
\hline \noalign{\smallskip}
BM25      & 0.153 & 0.123 & 0.527 & 0.173 \\            
PageRank    & 0.077 & 0.085 & 0.274 & 0.042 \\                             
QueryInURL  & 0.154 & 0.185 & 0.700 & 0.202 \\
RF        & 0.870{$^\blacktriangle$} & 0.736{$^\blacktriangle$} & 0.754{$^\blacktriangle$} & 0.804{$^\blacktriangle$} \\    
\hline \noalign{\smallskip}
\end{tabular}
\captionsetup{justification=centering,margin=1cm}
\caption{Retrieval performance using Bing ranks as ground truth (RF-B). Symbol $^\blacktriangle$ indicates significant improvement over BM25.}
\label{tbl:result1}
\end{table}

\begin{table}[htb]
\centering
\begin{tabular}{@{}ccccc@{}}
\toprule
      & P@1     &  P@10    & NDCG@10 & MAP\\
\hline \noalign{\smallskip}
BM25      & 0.385 & 0.269 & 0.517 & 0.270 \\            
PageRank    & 0.077 & 0.277 & 0.508 & 0.261 \\                             
QueryInURL  & 0.384 & 0.353 & 0.553 & 0.332 \\
RF        & 0.846{$^\blacktriangle$} & 0.769{$^\blacktriangle$} & 0.760{$^\blacktriangle$} & 0.798{$^\blacktriangle$} \\    
\hline \noalign{\smallskip}
\end{tabular}
\captionsetup{justification=centering,margin=1cm}
\caption{Retrieval Performance based on manual labels (RF-M). Symbol $^\blacktriangle$ indicates significant improvement over BM25.}
\label{tbl:result2}
\end{table}

Table \ref{tbl:result1} shows the performance of the baselines and of
our learning model RF-B, using Bing ranks as soft labels. Table
\ref{tbl:result2} shows the performance of the baselines and of our
learning model RF-M, using manual labels. In both settings, PageRank
performs poorly, mainly because it does not target directly the
query-driven relevance of the documents. BM25 has better performance,
and works better when judged by the manual labels than by comparing
with Bing ranks. Given that this baseline is the same in both
experiments, it reveals that our human relevance feedback is more
relaxed compared to the Bing soft labels. A surprising result is that
its performance is comparable to QueryInURL, and actually slightly
worse when we expand the cut-off threshold (from top-$1$ to top-$10$
results). In all three metrics Precision, NDCG and MAP. In other
words, for general entity search scenario with no special information
need, it might be more useful to look into the URL string and guess
whether a web page is a good resource to explore, than to examine the
raw anchor texts without any other context.

Applying learning models introduces a significant improvement on the
retrieval in both settings (both with $p$-value $< 0.001$). Note
however, that RF-M and RF-B values are not directly comparable, as
they are evaluated against different ground truths. In addition, the
high values are caused by our small samples which include many
positive search results, so cannot be generalized to performance of
our ranking model on our larger dataset A. We will investigate in
future work how the two models perform on substantially larger scale
datasets.

We also see that the NDCG scores of both models are lower than
Precision at the same cut-off threshold. This suggests that although we
are able to distinguish relevant and important documents from less
relevant ones, the ranks in the top results are still not good as they
should be (i.e. relevant but not important), leaving additional room
for future improvement of the ranking models.

\subsection{Feature Analysis}

\begin{table}[h]
\footnotesize
  \centering
    \begin{tabular}{ll}
    \toprule
    \multicolumn{1}{c}{\textbf{RF-M}} & \multicolumn{1}{c}{\textbf{RF-B}} \\
    \midrule
    1. InlinkCount & 1. Revision \\
    2. QueryInURL & 2. InlinkCount \\ 
    3. Revision & 3. DomainSize \\
    4. Idf & 4. QueryInURL \\
    5. DomainSize & 5. URLDepth \\
    6. PageRankDomain & 6. Idf \\
    7. PageRankCore & 7. AnchorTimeSpans \\
    8. LuceneTf & 8. FieldNorm \\
    9. AnchorTimeSpans & 9. LuceneTf \\
   10. AnchorFreq & 10. PageRankDomain \\
   11. URLDepth & 11. AnchorFreq \\ 
    \bottomrule
    
    \end{tabular}%
  \captionsetup{justification=centering}
  \caption{Top Features by Information Gain}
  \label{tbl:feature-analysis}%
\end{table}%

In Table \ref{tbl:feature-analysis}, we show the top features ordered
by their importance, as measured using Information Gain, of the two
learning models. InlinkCount and Revision are among the most
discriminating features. QueryInURL is also a very useful feature,
ranked second in RF-M and 4th in RF-B. In our experiments, anchor text
did not get into the top ranks, but stays at position 10th and 11th
for the two models. If we incorporate this feature with time
dimension, it becomes more discriminating (position 9 in RF-M and 7 in
RF-B). In addition, in our RF-B model (i.e. for distinguishing Bing
search results against others), except InLinkCount, all top-5 features
are ``light-weight'' features such as Revision, QueryInURL, URLDepth
which can be extracted easily from document metadata. For the
DomainSize feature, we can also perform a one-time process job to
compute the
values. 
This will help us to scale to larger subsets of Web archives.

\begin{table*}[htb]
\footnotesize
  \centering
    \renewcommand{\arraystretch}{1.4}
    \begin{tabular}{lll}
    \toprule
    \textbf{Query} & \multicolumn{1}{c}{\textbf{RF-M}} & \multicolumn{1}{c}{\textbf{RF-B}} \\
    \hline
    \bottomrule
    \multirow{5}{*}{\parbox{2cm}{Angela Merkel }} & \multicolumn{1}{|l}{\parbox{7.2cm}{\texttt{wiwo.de/koepfe-der-wirtschaft/angela-merkel/5288044.html}}} & \multicolumn{1}{|l}{\parbox{7.2cm}{\texttt{spiegel.de/international/topic/angela\_merkel}}}\\
    \cline{2-3}
     & \multicolumn{1}{|l}{\parbox{7.2cm}{\texttt{spiegel.de/international/topic/angela\_merkel}}} & \multicolumn{1}{|l}{\parbox{7.2cm}{\texttt{spiegel.de/thema/angela\_merkel}}}\\
    \cline{2-3}
     & \multicolumn{1}{|l}{\parbox{7.2cm}{\texttt{spiegel.de/thema/angela\_merkel}}} & \multicolumn{1}{|l}{\parbox{7.2cm}{\texttt{sueddeutsche.de/thema/angela\_merkel}}} \\
    \cline{2-3}
     & \multicolumn{1}{|l}{\parbox{7.2cm}{\texttt{sueddeutsche.de/thema/angela\_merkel}}} & \multicolumn{1}{|l}{\parbox{7.2cm}{\texttt{angela-merkel.de/}}} \\
    \cline{2-3}
    & \multicolumn{1}{|l}{\parbox{7.2cm}{\texttt{zeit.de/schlagworte/personen/angela-merkel/index}$^{\ast\ast}$}} & \multicolumn{1}{|l}{\parbox{7.2cm}{\texttt{welt.de/themen/angela-merkel/2}}}\\
    \bottomrule
    
    \multirow{5}{*}{\parbox{2cm}{Albert\\ Einstein }} & \multicolumn{1}{|l}{\parbox{7.2cm}{\texttt{spiegel.de/thema/albert\_einstein}}} & \multicolumn{1}{|l}{\parbox{7.2cm}{\texttt{spiegel.de/thema/albert\_einstein}}}\\
    \cline{2-3}
     & \multicolumn{1}{|l}{\parbox{7.2cm}{\texttt{amazon.de/albert-einstein-biographie-suhrkamp-taschenbuch/dp/3518389904}}} & \multicolumn{1}{|l}{\parbox{7.2cm}{\texttt{zitate-online.de/autor/einstein-albert}}}\\
    \cline{2-3}
     & \multicolumn{1}{|l}{\parbox{7.2cm}{\texttt{zitate-online.de/autor/einstein-albert}}} & \multicolumn{1}{|l}{\parbox{7.2cm}{\texttt{einsteingalerie.de/}}} \\
    \cline{2-3}
     & \multicolumn{1}{|l}{\parbox{7.2cm}{\texttt{sueddeutsche.de/thema/albert\_einstein}}} & \multicolumn{1}{|l}{\parbox{7.2cm}{\texttt{einstein-website.de/z\_kids/biographiekids.htm}$^{\ast}$}} \\
    \cline{2-3}
    & \multicolumn{1}{|l}{\parbox{7.2cm}{\texttt{einstein-gymnasium-hameln.de/schule/einstein/einstein.php}}} & \multicolumn{1}{|l}{\parbox{7.2cm}{\texttt{helles-koepfchen.de/albert\_einstein}}}\\
    \bottomrule

    \multirow{5}{*}{\parbox{2cm}{Volkswagen }} & \multicolumn{1}{|l}{\parbox{7.2cm}{\texttt{volkswagen.de/de.html}}} & \multicolumn{1}{|l}{\parbox{7.2cm}{\texttt{volkswagen.de/de.html}}}\\
    \cline{2-3}
     & \multicolumn{1}{|l}{\parbox{7.2cm}{\texttt{autohaus24.de/neuwagen-kaufen/volkswagen}}} & \multicolumn{1}{|l}{\parbox{7.2cm}{\texttt{auto.de/kfzkatalog/vw}}}\\
    \cline{2-3}
     & \multicolumn{1}{|l}{\parbox{7.2cm}{\texttt{volkswagen-nutzfahrzeuge.de/de.html}}} & \multicolumn{1}{|l}{\parbox{7.2cm}{\texttt{motor-talk.de/forum/volkswagen-b22.html}}} \\
    \cline{2-3}
     & \multicolumn{1}{|l}{\parbox{7.2cm}{\texttt{motor-talk.de/forum/volkswagen-b22.html}}} & \multicolumn{1}{|l}{\parbox{7.2cm}{\texttt{volkswagen-nutzfahrzeuge.de/de.html}}} \\
    \cline{2-3}
    & \multicolumn{1}{|l}{\parbox{7.2cm}{\texttt{spiegel.de/thema/vw}}} & \multicolumn{1}{|l}{\parbox{7.2cm}{\texttt{autohaus24.de/neuwagen-kaufen/volkswagen}}}\\
    \bottomrule

    \multirow{5}{*}{\parbox{2cm}{Bruce\\ Willis}} & \multicolumn{1}{|l}{\parbox{7.2cm}{\texttt{sueddeutsche.de/thema/bruce\_willis}}} & \multicolumn{1}{|l}{\parbox{7.2cm}{\texttt{sueddeutsche.de/thema/bruce\_willis}}}\\
    \cline{2-3}
     & \multicolumn{1}{|l}{\parbox{7.2cm}{\texttt{amazon.de/bruce-willis/e/b000apunwa}$^{\ast}$}} & \multicolumn{1}{|l}{\parbox{7.2cm}{\texttt{kino.de/star/bruce-willis/8453}$^{\ast\ast}$}}\\
    \cline{2-3}
     & \multicolumn{1}{|l}{\parbox{7.2cm}{\texttt{new-video.de/darsteller-bruce-willis}}} & \multicolumn{1}{|l}{\parbox{7.2cm}{\texttt{filmstarts.de/personen/197-bruce-willis.html}}} \\
    \cline{2-3}
     & \multicolumn{1}{|l}{\parbox{7.2cm}{\texttt{kino.de/star/bruce-willis/8453}$^{\ast\ast}$}} & \multicolumn{1}{|l}{\parbox{7.2cm}{\texttt{moviemaze.de/celebs/15/1.html}$^{\ast\ast}$}} \\
    \cline{2-3}
    & \multicolumn{1}{|l}{\parbox{7.2cm}{\texttt{moviemaze.de/celebs/15/1.html}$^{\ast\ast}$}} & \multicolumn{1}{|l}{\parbox{7.2cm}{\texttt{new-video.de/darsteller-bruce-willis}}}\\
    \bottomrule
    
    \end{tabular}%
  \captionsetup{justification=centering}
  \caption{Comparison of top-$5$ results between RF models learnt using using manual labels and Bing rank. $\ast$ indicates URLs that are not alive anymore in the Web, whereas $\ast\ast$ indicates URLs that are now redirected ones, checked in April 2016.}
  \label{table:anecdotal}%
\end{table*}%

\subsection{Case Study Analysis}
\label{sec:ex}

To look at our results in more detail, in Table \ref{table:anecdotal}
we show the comparison of top-$5$ results as retrieved by the two
models. To guarantee a fair comparison, we only retrieve the documents
that are available in the results of both models, and show them in the
same order as they are ranked in each result list.


\vpar
\noindent \textit{Angela Merkel}: The top-$1$ result using the model
learnt by manual labels is the biography web page of Angela Merkel
hosted by \texttt{wiwo.de}, not included at top rank for the Bing
experiment. The pages at the 5th rank of both models are news topic
pages, although the first one (\texttt{zeit.de} page) is no longer
available in the actual Web, and is replaced by another page on the
same topic.

\vpar
\noindent \textit{Albert Einstein}: The first-ranked document is the
same for both models. For the other 4 top results, however, the model
learnt by manual labels tends to be more diverse, with one result
(\texttt{einstein-gymnasium-hameln.de}) having nothing to do with
Albert Einstein as the physicist. In constrast, the model using Bing
ranks gives better results, all biographic information to the
physicist Albert Einstein. Also, it manages to find the relevant page
\texttt{einstein-website.de/z\_kids/biographiekids.htm} in the
archive, which is no longer available on the Web. Looking into the
user evaluation feedback, we see that users tend not to agree on
marginal cases where Web pages about Albert Einstein actually discuss
some distantly related information such as the places named after him,
or Web sites simply using the name Albert Einstein to symbolize
outstanding intelligence, so our manual labels are not really very
reliable in this case.

\vpar
\noindent \textit{Volkswagen}: Both models give good results. Although
the ranks of documents differ, it is difficult to judge which one is
really better than the other. We believe the reason is that the
underlying information need is very diverse: User can search for the
company, or search for the car products, either for commercial
purposes or to look up technical instructions. This ambiguity makes it
difficult for both models to learn a sensible meaning from simple
labeling approaches.

\vpar
\noindent \textit{Bruce Willis}: Both models agree on the first
position, but RF-M gives second position to an Amazon page with a list
of Bruce Willis' films, which is now only available in
\texttt{archive.org}. It also pushes redirect URLs (fourth and fifth
positions), because of the anchor texts pointing to. Using Bing rank,
RF-B delivers better quality, with all top 5 results biographic and
topic pages about the movie star. Looking closely to the results, we
can see that most of the top results are about movies where Bruce
Willis acted and not about his information as a celebrity. Perhaps
such information need (search for Bruce Willis' films) is more
popular, as Bruce Willis is not known for many scandals and thus
solicits less personal comments on the Web than other
celebrities. This bias is handled better by relying on Bing labeling,
rather than a small number of users labeling results.

\subsection{Discussion} 

In general, for top results, both models agree in many cases, with some
small difference between queries such as ``\textsf{Angela
  Merkel}''. Comparing the analysis in Section \ref{sec:ex} with our
empirical evaluation in Section \ref{sec:res1}, we see that in several
cases the model learnt using Bing ranks as soft labels tends to prefer
documents which are more focused on the entity and also more
important. 
As the results of Bing come from advanced models with a much richer
set of dimensions and features, the integration of more features into
our model will certainly enable it to reflect better the main aspects
embedded in current search engine rankings.  Given the small size of
samples in our evaluation and the inclusion of all ``good'' results
(from Bing), it would be interesting to see how good our models can
deal with larger subsets of the archive, which will be more noisy and
include more spamming page.

For entity queries, we see that some highly valuable pages, such as
Wikipedia pages for the biographical information of the entities, are
not shown in the top results. It is interesting to see how we can
encode such credits with more features, tailored to entity exploration
scenarios. In this context, other issues such as spamming URLs, user
qualitative evaluation, etc. should be analysed deeper. For instance,
the comparison between two models should not be limited to
quantitative assessment, but also be extended to the user satisfaction
measurement, to really understand the usefulness of the document
non-content evidences. We can investigate further the
search scenarios for different types of entities (e.g., the
information need when user searches for a celebrity is substantially
different than when she searches for a scientist, or a travelling
location), and design different per-type questions for user
experiments.

\section{Conclusion}
\label{sec:conclusion}




In this paper, we conducted a first study about the influence of
non-content evidences on searching in Web archives. By using Bing
search results as proxy for relevance, and by focusing our search
scenario on exploring unambiguous entities, we can identify the
correlation between different features and the quality of
documents. We find out that although anchor texts are useful resources
for searching in the Web, applying them to Web archives require more
thoughts, taking into account other evidences such as entity types,
URL, etc.

We also study the possibility of combing multiple evidences to improve
the ranking. The result is promising, and can be a good starting point
for an efficient ranking framework that does not rely on full content
indexing of documents, at least for an entity exploration scenario,
where the user has no special ad-hoc information need.

There are several other directions to extend this work. For instance,
we aim to target different search scenarios in the Web archive, such as navigational search for particular (but partially forgotten)
resources in the past. We also aim to improve the retrieval component
of our work, to facilitate both ambiguous and unambiguous queries, and
to scale and evaluate our results for larger subsets of Web archives.

\vspace{0.25cm}
\noindent\textbf{Acknowledgements.} {This work was funded by the 
ERC Advanced Grant ALEXANDRIA (under the grant No. 339233). We thank the reviewers for the fruitful discussion and suggestions on future work. We also thank our colleague Philipp Kemkes for setting up the usage of Bing search API to facilitate our experiments.
}

\bibliographystyle{abbrv}
\vspace{-0.2cm}
\begin{small}

\end{small}

\end{document}